\documentclass[preprint,amsmath,amssymb,aps,showkeys,showpacs]{revtex4}

\usepackage[colorlinks=true, pdfstartview=FitV, linkcolor=blue, citecolor=red, urlcolor=magenta]{hyperref}
\usepackage{graphicx}
\usepackage{latexsym}
\usepackage{amsmath}
\usepackage{amsfonts}
\usepackage{amssymb}
\usepackage{verbatim}
\usepackage{dcolumn}
\usepackage{amssymb}
\usepackage{bm}
\usepackage{float}
\usepackage{subfigure}

\usepackage[english]{babel}
\graphicspath{{Figures/}}

\begin{document}

\newcommand{\NITK}{
\affiliation{Department of Physics, National Institute of Technology Karnataka, Surathkal  575025, India.}
}
\newcommand{\TKM}{\affiliation{
Department of Physics, T.K.M College of Arts and Science, Kollam, Kerala 691005, India.}
}
\newcommand{\UW}{\affiliation{
Department of Oral Health Sciences, School of Dentistry, University of Washington, Seattle, WA 98195, USA.
}}
\newcommand{\gk}{\affiliation{Department of Physics, Government College, Kasaragod, Kerala, 671123, India.}}
\title{Dynamic Phase Transition of Black Holes in Massive Gravity}

\author{T. K. Safir}
\email{stkphy@gmail.com}
\NITK
\TKM
\author{A. Naveena Kumara}
\email{naviphysics@gmail.com}
\NITK
\author{Shreyas Punacha}
\email{shreyas4@uw.edu}
\UW
\author{C. L. Ahmed Rizwan}
\email{ahmedrizwancl@gmail.com}
\gk
\author{C. Fairoos}
\email{fairoos.phy@gmail.com}
\TKM
\author{Deepak Vaid}
\email{dvaid79@gmail.com}
\NITK

\begin{abstract}
The dynamical properties of small-large black hole phase transition in dRGT non-linear massive gravity theory are studied based on the underlying free energy landscape. The free energy landscape is constructed by specifying the Gibbs free energy to every state, and the free energy profile is used to study the different black hole phases. The small-large black hole states are characterized by probability distribution functions and the kinetics of phase transition are described by the Fokker-Planck equation. Further, a detailed study of the first passage process is presented which describes the dynamics of phase transitions. Finally, we have investigated the effect of mass and topology on the dynamical properties of phase transitions of black holes in dRGT non-linear massive gravity theory.
\end{abstract}


\maketitle


\section{Introduction}
\label{sec1}

A black hole is, hopefully, an excellent tool that could be used to unravel the nature of gravity at the quantum mechanical level. The well-established connection between a black hole and an ordinary thermodynamic system \citep{Hawking:1974sw, Bekenstein1973} has laid a strong foundation for this prospect. The hope is to understand the underlying microscopic description corresponding to the macroscopic thermodynamic features of black holes. Consequently, rigorous studies were conducted to understand gravity from the viewpoint of thermodynamics and statistical physics, and a lot of advancements have been made in recent years. One of the profound results in this regard is due to Hawking and Page who demonstrated the existence of phase transition in the phase space of an Anti-de Sitter (AdS) Schwarzschild black hole \citep{Hawking:1982dh}. Later, a charged black hole in AdS spacetime is shown to exhibit a first-order phase transition similar to a van der Waal liquid-gas system \citep{Chamblin:1999tk, Chamblin:1999hg}. These calculations are carried out in an extended phase space where the cosmological constant ($\Lambda$) is treated as thermodynamic pressure and its conjugate quantity as a thermodynamic volume. It turns out that various class of black hole solutions in asymptotically AdS spacetime possess rich phase structure and shows critical phenomena and phase transitions \citep{Kubiznak2012, Gunasekaran2012, Kubiznak:2016qmn, Cai:2013qga,Wei:2013fs, Zou:2014sr, Rajagopal:2014ewa, Mo:2014qsa, Xu:2015rfa, Dayyani:2019ys, Dayyani:2017fuz}.\\

The microscopic degrees of freedom responsible for the thermodynamic quantities of a van der Waal system are well known. However, a satisfactory explanation of black hole microstructures is yet to be attained. There are many interesting studies that prob black hole microstructures using various techniques in geometrothermodynamics \citep{Wei:2015iwa, Wei:2019uqg, Wei:2019yvs}. Recently, a method based on the free energy landscape has been proposed in which a black hole is considered as the macroscopic emergent state of microscopic degrees of freedom corresponding to the space-time \citep{Li:2020khm}. In this formalism, the black hole states are described by Gibbs free energy in terms of an order parameter. The order parameter is realized as a coarse-grained description that reflects the features of the microscopic degrees of freedom of the system. The free energy distributed among the state space constitutes a free energy landscape and the formalism is used to study the phase transitions in many systems\citep{Frauenfelder:1991fs, Goldenfeld:1992qy, Wang2015LandscapeAF}. In the free energy landscape construction for black hole systems, the horizon radius is taken as the order parameter. The kinetics of black hole phase transitions are then studied using the probabilistic Fokker-Planck equation on the free energy landscape. This methodology has been applied to study the phase transitions in Einstein's gravity and massive gravity theories\citep{Li:2020khm}. The calculation can be extended by solving the Fokker-Planck equation subjected to suitable boundary conditions to obtain the stationary distributions of black hole states at different temperatures. Further, one can explore the first passage time of the probabilistic evolution between black hole states. The dynamics and kinetics of black hole phase transitions on free energy landscape are investigated for five-dimensional Gauss-Bonnet \citep{Wei:2020rcd} and Reissner-Nordstrom  \citep{Li:2020nsy} black holes in AdS spacetime.\\

The theory of general relativity (GR) describes how gravity works in four dimensions and is based on massless gravitons with two degrees of freedom. Even though GR is a remarkable theory, it suffers certain fundamental issues\citep{Capozziello:2011et}. Naturally, many attempts have been made by modifying Einstein-Hilbert action.
 The massive gravity theory was formulated as a straightforward modification of general relativity in which the graviton acquires a non-zero mass. The advantage of such a theory is that it explains the accelerated expansion of our universe without introducing a bare cosmological constant. The initial massive gravity model was proposed by Fierz and Pauli in 1939 \cite{Fierz:1939ix}. However, the theory did not produce correct GR limits in the massless case. Also, the non-linear modifications of the initial theory lead to ``Boulware-Deser" ghost instability \citep{PhysRevD.6.3368}. Later, de Rham, Gabadadze, and Tolley (dRGT) proposed a  ``Boulware-Deser" ghost-free theory with proper GR limit by adding higher-order interaction terms to the Einstein-Hilbert action\citep{PhysRevLett.106.231101}.  Consequently, the black hole solutions in this modified gravity theory and their thermodynamic characteristics were extensively investigated. The van der Waal-like features and other related topics such as triple point, Reentrant phase transitions, heat engines, and throttling process were also studied \cite{Cai:2014znn, Hendi:2017fxp, Hendi:2015bna, Mirza:2014xxa, Fernando:2016qhq,  Ning:2016usb, Zou:2016sab, Liu:2018jld, Hendi:2017bys, Yerra:2020bfx}. In addition, several methods to probe the microstructure of black hole solutions in massive gravity were  presented using various thermodynamic-geometry approaches \cite{Chabab:2019mlu, Wu:2020fij, Yerra:2020oph, Safir:2022vkf, Fairoos:2023jvw}.\\

This paper focuses on the dynamical properties of black hole phase transitions in dRGT massive gravity theory using the formalism of free energy landscape. The organization of the paper is as follows. In section \ref{first} and \ref{sec3}, we discuss the thermodynamic structure and phase transition of black hole solutions in dRGT non-linear massive gravity in terms of free energy. In section \ref{sec4}, we study the Fokker-Planck equation, and the numerical solutions are obtained for reflecting boundary conditions. Also, we discuss the numerical results of the first passage time leaking from a small black hole to large black hole phases. In section \ref{sec5}, we investigate the effect of mass and topology on the dynamical properties of the small-large black hole phase transition in the massive gravity theory. The results are summarised in section \ref{sec6}.


\section{Thermodynamic Characterisation and phase transition}\label{first}

We briefly discuss the spacetime structure and the thermodynamic properties of black hole solutions of four-dimensional dRGT non-linear massive gravity theory in AdS space. The action for the Einstein-dRGT gravity coupled to a non-linear electromagnetic field is obtained by adding interaction terms to the Einstein-Hilbert Lagrangian as\citep{Vegh:2013sk},
\begin{equation}
S=\int d^4 x \sqrt{-g} \left[ \frac{1}{16 \pi}\left[ R+\frac{6}{L^2}+m^2\sum_{i=1}^4 c_i \ \mathcal{U}_i(g,f)\right ]-\frac{1}{4\pi } F_{\mu \nu}F^{\mu \nu}\right],
\end{equation} 
 where $F_{\mu \nu}=\partial_\mu A_\nu-\partial_\nu A_\mu $ is the electromagnetic field tensor with vector potential $A_\mu$, $L$ is AdS radius which is related to the cosmological constant $(\Lambda)$ by $L^2=-3/\Lambda$, $m$ is related to the graviton mass, and $c_i$ are coupling parameters. Further, $f_{\mu \nu}$ is a symmetric tensor known as reference metric coupled to the space-time metric $g_{\mu \nu}$. Graviton interaction terms are represented by symmetric polynomials $\mathcal{U}_i$, and are obtained from a $4 \times 4 $ matrix $\mathcal{K}^\mu_\nu =\sqrt{g^{\mu \alpha} f_{\nu \alpha}}$, which have the following forms,
\begin{eqnarray*}
\begin{split}
\mathcal{U}_1 &= [\mathcal{K}],\\
\mathcal{U}_2 & = [\mathcal{K}]^2-[\mathcal{K}^2],\\
\mathcal{U}_3 & = [\mathcal{K}]^3-3[\mathcal{K}^2][\mathcal{K}]+2[\mathcal{K}^3],\\
\mathcal{U}_4 & = [\mathcal{K}]^4-6[\mathcal{K}^2][\mathcal{K}]^2+8[\mathcal{K}^3][\mathcal{K}]+3[\mathcal{K}^2]^2-6[\mathcal{K}^4].
\end{split}
\end{eqnarray*}
The black hole solutions to the above action can be obtained in the following form \citep{Cai:2014znn, PhysRevD.95.021501}:

\begin{equation}
ds^2=-f(r)dt^2+\frac{1}{f(r)}dr^2+r^2h_{ij} dx_i dx_j,
\end{equation}
where $h_{ij}$ is the metric on the two-dimensional hypersurface. The topological parameter ($k$) can take values $0,-1$, or $1$, representing planar, hyperbolic, and spherical topology, respectively. Choosing the reference metric $f_{\mu\nu}=\text{diag}(0,0,c_0^2 \ h_{ij})$, the values of the interaction terms $\left(\mathcal{U}_i\right)$  become $\mathcal{U}_1=\frac{2c_0}{r}$,~~ $\mathcal{U}_2=\frac{2c_0^2}{r^2},~~\mathcal{U}_3= \mathcal{U}_4= 0$ \citep{Cai:2014znn}. Now, the metric function reduces to,
\begin{equation}\label{metric_1}
f(r)=k-\frac{m_0}{r}-\frac{\Lambda r^2}{3} +\frac{q^2}{r^2}+m^2\left( \frac{c_0 c_1}{2}r +c_0^2 c_2\right).
\end{equation}
Here, the parameters $m_0$ and $q$ are related to the black hole mass and charge respectively. The graviton mass is represented by $m$. Note that in the limit $m \rightarrow 0$, the metric function reduces to Reissner- Nordstrom black hole solution in AdS space. Now, the event horizon ($r_+$) is obtained by solving the equation $f(r_+)=0$.  The Hawking temperature of the black hole is related to its surface gravity by the relation $T_H = \kappa/({2\pi})$, where the surface gravity $\kappa = f'(r_+)/2$. As in the case of an asymptotically AdS black hole in four dimensions, one can relate the thermodynamic pressure with cosmological constant \citep{Kastor:2009wy, Dolan:2010ha, Dolan:2011xt} as,
\begin{equation*}
P = -\frac{\Lambda}{8\pi}.
\end{equation*}
At this point, the mass, temperature, and entropy of the black hole in Einstein-dRGT gravity coupled to a non-linear electromagnetic field can be expressed in terms of the horizon radius and pressure as follows:
\begin{eqnarray*}
M&=& \left( \frac{r_+}{2}(k +c_0^2 c_2 m^2)+\frac{c_0c_1 m^2 r_+}{2	}+\frac{8}{3}\pi P r_+^2 +\frac{q^2}{4r_+^2}\right),\\
T_H&=& \left( 2P r_++\frac{k+c_0^2c_2m^2}{4\pi r_+	}-\frac{q^2}{16\pi r_+^3} +\frac{c_0c_1 m^2}{4\pi}\right),\\ 
S&=& \pi r_+^2.
\end{eqnarray*}
Thermodynamic volume is obtained as,
\begin{equation}
V=\frac{4}{3}\pi r_+^3 =\frac{\pi}{6}v^3,
\end{equation} 
where $v=2r_+$ is the specific volume. The equation of state of the system $P=P(T_H, v)$ is,
\begin{equation}
P=\frac{q^2}{2\pi v^4}-\frac{k+c_0^2c_2m^2}{2\pi v^2}+\frac{T_H}{v}-\frac{c_0 c_1 m^2}{4\pi v}. 
\end{equation}
 This expression indicates that black holes in massive gravity theory exhibit vdW fluid-like behavior. Now, the critical points corresponding to the first order phase-transition between a large black hole phase (LBH) and a small black hole phase (SBH) in the extended phase space is obtained from the below conditions.
\begin{equation}
\left( \frac{\partial P}{\partial v} \right)_{T_H=0}, \qquad \text{and} \qquad \left( \frac{\partial ^2 P}{\partial v^2} \right)_{T_H}=0.
\end{equation}
Accordingly, we obtain,
\begin{equation}
P_c=\frac{(k+m^2c_2c_0^2)^2}{24\pi q^2}. \qquad
\end{equation} 
\begin{figure*}[tbh]
\centering
\subfigure[ref2][]{\includegraphics[scale=0.5]{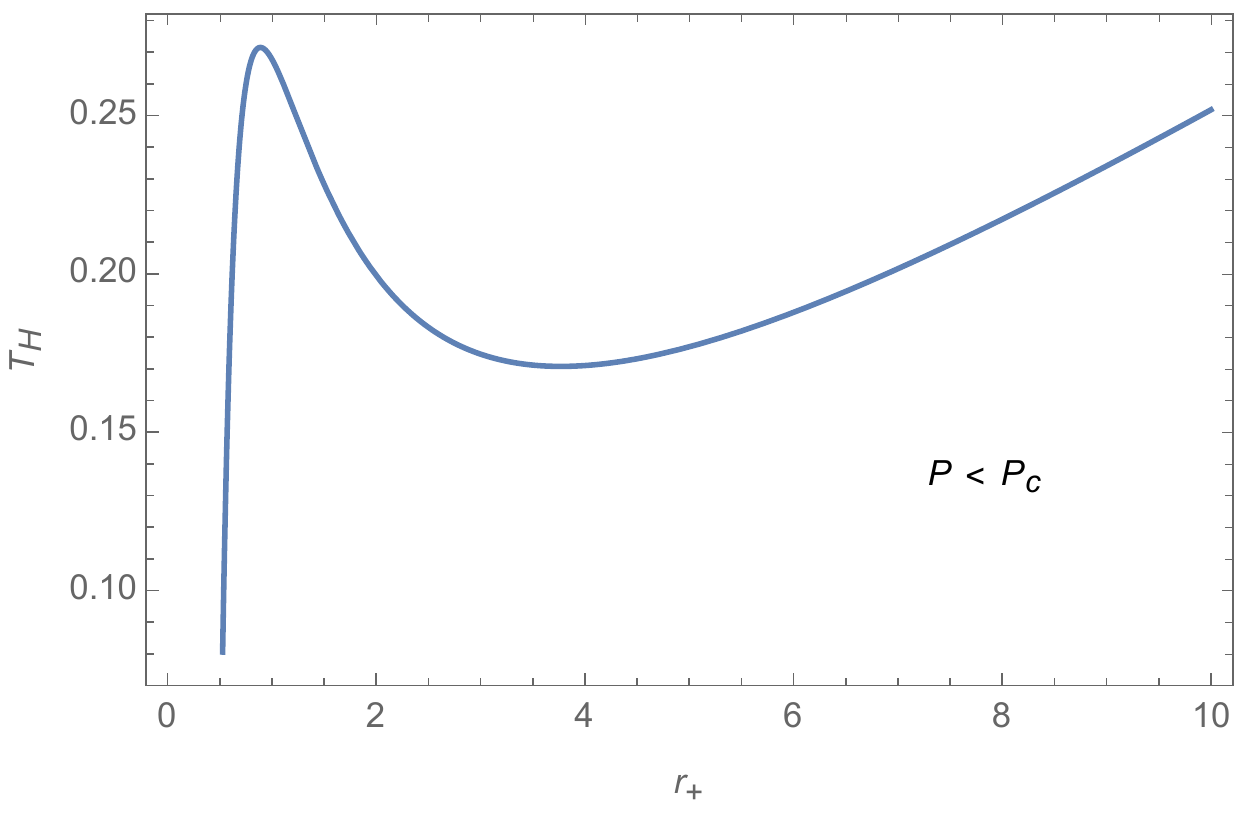}
\label{Th_r_1}}
\qquad
\subfigure[ref3][]{\includegraphics[scale=0.5]{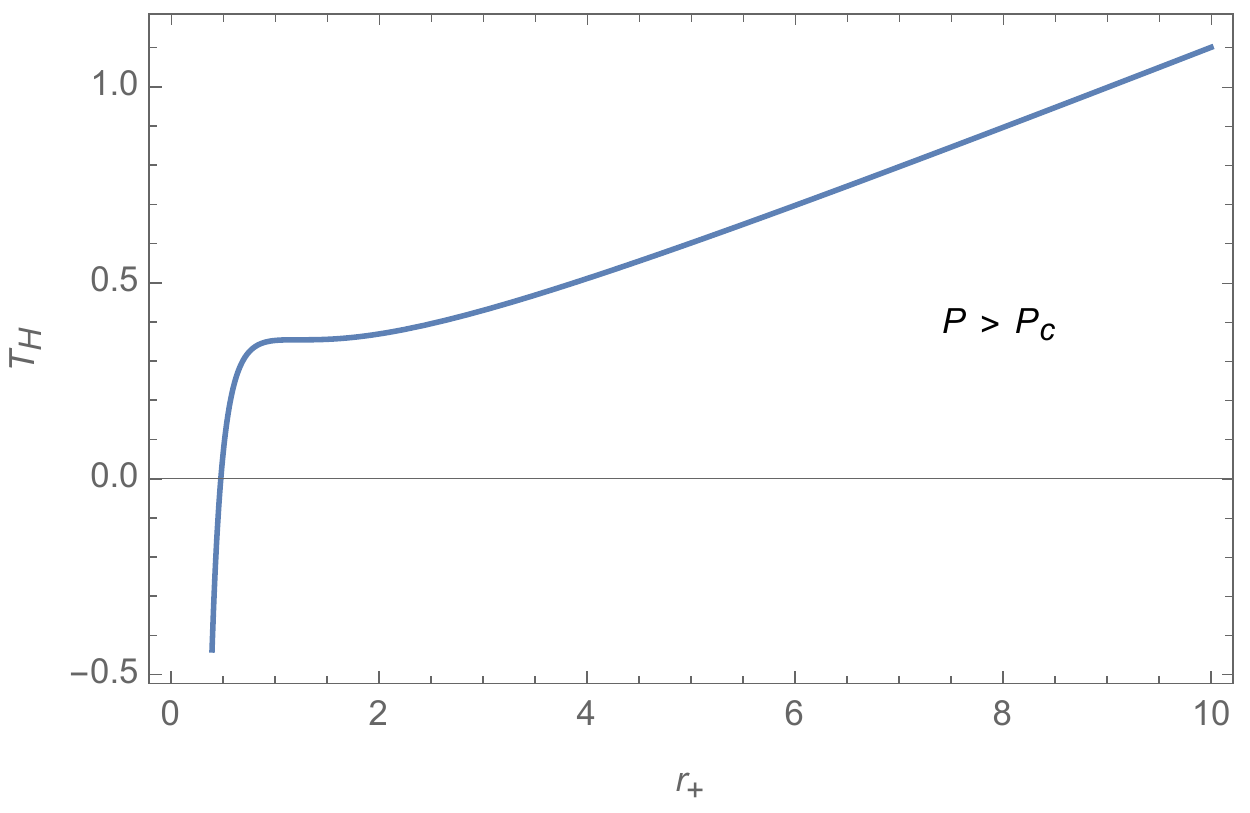}
\label{Th_r_2}}
\caption{A plot showing black hole temperature as a function of $r_+$ (a) when $P<P_c$  and (b) $P>P_c$.}
\label{Criticalpressure}
\end{figure*}

The black hole temperature is a monotonic function of $r_+$ for $P>P_C$. Whereas below the critical pressure, $T_H$ have local minimum and local maximum values. Note that this characteristic is independent of the topology of the system. In Fig. [\ref{Criticalpressure}] we have depicted a typical behaviour of $T_H$ as a function of $r_+$ for both $P<P_c$ and $P>P_c$.\\

The local minima and local maxima of black hole temperature for $P<P_c$ are determined by,
\begin{equation*}
\frac{\partial T_H}{\partial r_+} = 0,
\end{equation*}
giving the solutions,
\begin{equation*}
r_{min/max} = \frac{1}{4 \sqrt{\pi}}\Big[ \frac{k+c_0^2 m^2 c_2 \mp \left(k+c_0^2m^2c_2-24 \pi P q^2\right)^{\frac{1}{2}}}{P}\Big]^{\frac{1}{2}}.
\end{equation*}
The corresponding values of black hole temperature are given by,
\begin{eqnarray*}
T_{min/max} &=& \frac{1}{4\sqrt{\pi}}\Bigg[ c_0 c_1 m^2 \pi^{\frac{3}{2}} - \frac{16 \pi q^2}{\left( \frac{k+c_0^2 m^2 c_2 \mp \left(k+c_0^2m^2c_2-24 \pi P q^2\right)^{\frac{1}{2}}}{P}\right)^{\frac{3}{2}}} \\
& +& \frac{4\left(k+c_0^2m^2c_2\right)}{\left( \frac{k+c_0^2 m^2 c_2 \mp \left(k+c_0^2m^2c_2-24 \pi P q^2\right)^{\frac{1}{2}}}{P}\right)^{\frac{1}{2}}} +2 P \left( \frac{k+c_0^2 m^2 c_2 \mp \left(k+c_0^2m^2c_2-24 \pi P q^2\right)^{\frac{1}{2}}}{P}\right)^{\frac{1}{2}}\Bigg].
\end{eqnarray*}
Further analysis shows, when the black hole temperature lies $T_{\text{min}} < T_H < T_{\text{max}}$, there exist three branches of black hole solution (i.e. small, intermediate, and large black hole), in which the intermediate solution is unstable. Also, there is a first-order phase transition from the SBH to the LBH similar to the van der Waals liquid-gas system. As mentioned before, free energy landscape is a useful tool to study phase transitions. In the following section, we investigate the characteristics of first-order phase transition of black holes using Gibbs free energy landscape. 
\section{Gibbs free energy landscape}
\label{sec3}
\begin{figure*}[tbh]
\includegraphics[scale=0.6]{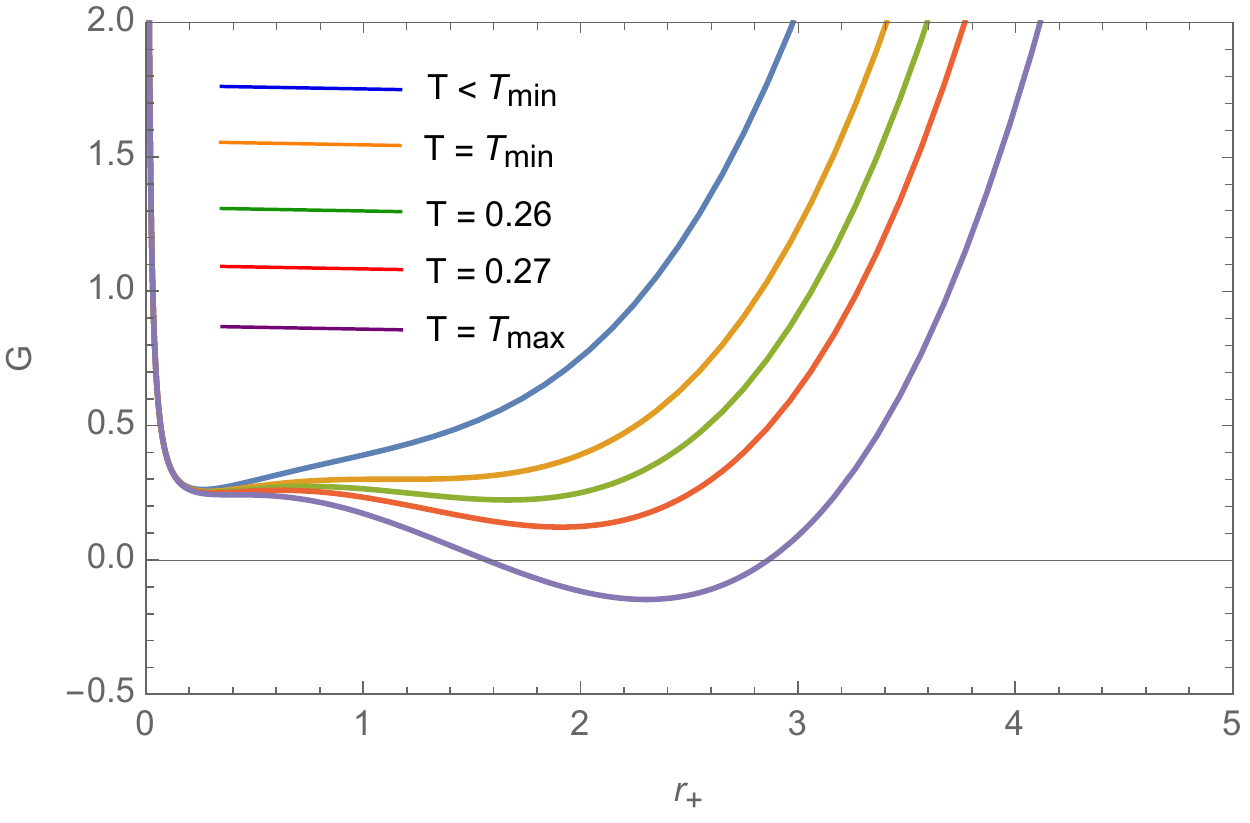}

\caption{\small{A plot of $G$ as a function of $r_+$ for $P<P_c$ for different temperatures. Here, $k=1, m=c_0=1, q=0.5, c_1=1, c_2=0.3, T_{max} = 0.29$, and $T_{min} = 0.24$}}
\label{G_vs_r}
\end{figure*}

 We consider a canonical ensemble composed of a series of black hole spacetimes at a given temperature $T$. The free energy landscape is constructed by specifying Gibbs free energy to every spacetime state. The stable branches of black hole solutions are associated with on-shell Gibbs free energy that can be obtained either from the Euclidean action \citep{Dolan:2011xt} or from the thermodynamic relationship,
 \begin{equation}\label{free_energy}
  G=M-T_H S. 
  \end{equation}
 However, we need an expression for a generalized free energy (off-shell) to describe other black hole states as well (such as transient or excited states that are not a solution to the gravity theory). According to the free energy landscape formalism, the generalized free energy is obtained by replacing $T_H$ with the ensemble temperature ($T$) in Eq. (\ref{free_energy}). 
\begin{equation}
G= M- T\ S = \frac{r_+}{2}\left(k+\frac{q^2}{r_+^2}+\frac{8}{3}P \pi r_+^2 + m^2\left(c_0^2 c_2+\frac{c_0c_1r_+}{2}\right)\right)-\pi T r_+^2.
\end{equation}
In this construction, the black hole radius is taken as the order parameter describing the microscopic degrees of freedom of the system. The Gibbs free energy landscape as a function of the black hole radius for $P<P_c$ at different values of the temperature is studied in Fig. [\ref{G_vs_r}]. When $T<T_{\text{min}}$, there is only one global minimum for the Gibbs free energy and corresponds to the pure radiation phase. At $T=T_{\text{min}}$, there is an inflection point. For $T > T_{\text{min}}$, two black hole phases emerge (small and large black hole phases). The smaller and larger black holes correspond to  local minima of Gibbs free energy.\\

For $T_{\text{min}} < T < T_{\text{max}}$, the Gibbs free energy has three local extremum values, in which two are stable and one is unstable. The extremum values are determined by,
\begin{eqnarray}\label{radius}
\frac{\partial G}{\partial r_+} &=& 0 \nonumber \\
 \Rightarrow \quad \frac{k}{2} &+& \frac{m^2}{2}\left( c_0^2 c_2+c_0 c_1 r_+\right) - \frac{q^2}{2r_+^2}-2 \pi r_+ \left(T-2 P r_+\right)=0.
 \end{eqnarray}
 Solving this equation for radius, we obtain the expressions for small, large, and intermediate black hole states. Out of three branches, the small and large states are locally thermodynamically stable (Gibbs free energy has a minimum). The Gibbs free energy corresponding to the intermediate black hole state has a maximum value and therefore is unstable.  Now, the expression for the free energy corresponding to these small, large, and intermediate states obtained in terms of their radius as,

 \begin{equation}
 G_{s/m/l} = \frac{r_{s/m/l} }{2}\eta+\frac{3q^2}{4r_{s/m/l} }-\frac{2}{3}P \pi r_{s/m/l} ^3,
  \end{equation}
  where $\eta = k+c_2m^2$. One can quantify the free energy landscape topography by constructing free energy barrier heights from SBH to the intermediate black hole ($G(r_m)-G(r_s)$) and from the intermediate black hole to LBH ($G(r_m)-G(r_l)$). We observe that both barrier heights are monotonic functions of temperature. Similar to the RN-AdS black holes, the barrier height from SBH to the intermediate black hole decreases with temperature, whereas, the barrier height from the intermediate black hole to the large black hole increases \cite{Li:2020nsy}.\\

\begin{figure*}[tbh]
\includegraphics[scale=0.6]{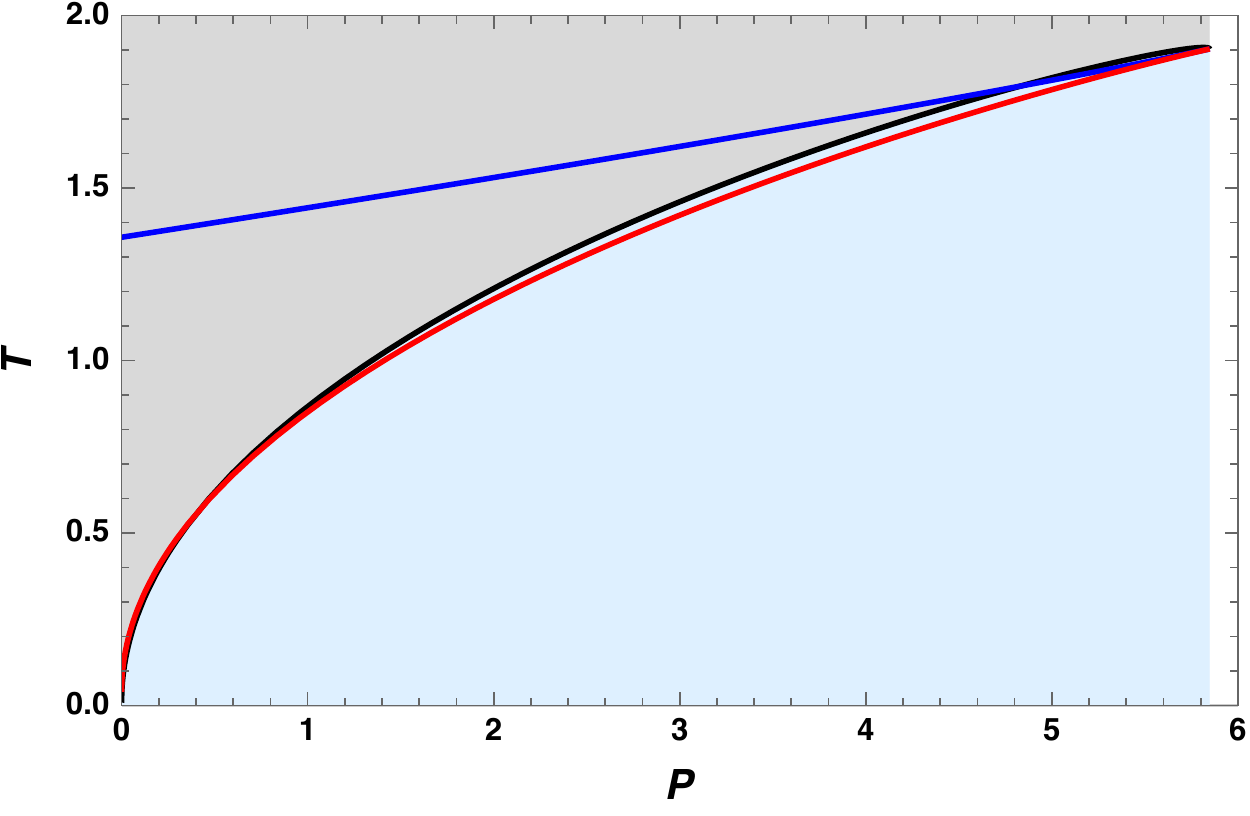}

\caption{\small{The phase diagram of dRGT black holes. Here, $T_{\text{max}}$, $T_{\text{min}}$, and $T_{\text{trans}}$ are plotted as a function of $P$ from 0 to $P_c$. SBH is stable in the blue region whereas LBH is stable in the grey region. In this plot, the black line represents the coexisting curve.}}
\label{Phase_diagram}
\end{figure*}

  As the temperature increases, the local minimum of Gibbs free energy lowers until it becomes zero at $T=T_{\text{trans}}$, where $T_{\text{trans}}$ is known as the transition temperature. At this point, the Gibbs free energy of large and small black holes are equal. Therefore, the transition temperature can be obtained from the following equations:
  \begin{eqnarray*}
  \frac{1}{2} \left(\eta+c_1 r_s\right) - \frac{q^2}{2r_s^2}-2 \pi r_s \left(T-2 P r_s\right)=0,\\
    \frac{1}{2} \left(\eta+c_1 r_l\right) - \frac{q^2}{2r_l^2}-2 \pi r_l \left(T-2 P r_l\right)=0,\\
    \text{and} \quad \frac{r_{s} }{2}\eta+\frac{3q^2}{4r_{s} }-\frac{2}{3}P \pi r_{s} ^3=\frac{r_{l} }{2}\eta+\frac{3q^2}{4r_{l} }-\frac{2}{3}P \pi r_{l} ^3;
  \end{eqnarray*}
  The expressions for the small and large black hole radii is readily obtained as,
  \begin{equation*}
  r_s = \frac{1}{8 \left(\pi P\right)^{\frac{3}{2}}}\Big[\left(\eta+\omega\right)\left(\eta\left(\eta+\omega\right)-12 P \pi q^2\right)\Big]^{\frac{1}{2}},
  \end{equation*}
  \begin{equation}
  r_l = \frac{1}{4 \pi P\left(\eta+\omega\right)}\Big[\eta\left(\eta+\omega\right)-12 P \pi q^2+4\sqrt{\pi}\left(P\left(\eta+\omega\right)\left(\eta\left(\eta+\omega\right)-12 P \pi q^2\right)\right)^{\frac{1}{2}}.
  \end{equation}
 Here, $\omega= \sqrt{\eta^2-24 P \pi q^2}$. Now the transition temperature is obtained as,
 \begin{small}
  \begin{eqnarray*}
  T_{\text{trans}} &=& \frac{6 \sqrt{P \pi} t^2 \left(\eta+\omega\right) - 2 q^2 \left(P \pi\right)^{\frac{3}{2}}\left(49 \eta +13 \omega\right) + c_1 m^2 \left(\eta + \omega\right)^{\frac{3}{2}} \sqrt{\eta\left(\eta+\omega\right)-12 \pi P q^2}
}{2 \pi \left(\eta+\omega\right)^{\frac{3}{2}} \sqrt{\eta\left(\eta+\omega\right)-12 \pi P q^2}}.
  \end{eqnarray*}
  \end{small}
A thermodynamic phase diagram is given by plotting $T_{\text{max}}$, $T_{\text{min}}$, and $T_{\text{trans}}$  as a function of $P$. For a chosen parameters of the system ($k=1, m=1, q=c_1=c_2=0.05$), the curves divide the $P-T$ plane into four thermodynamic phase regions. In Fig. [\ref{Phase_diagram}], the blue, black, and red lines represent $T_{\text{max}}$, $T_{\text{trans}}$, and $T_{\text{min}}$ respectively.\\
Note that the structure of the phase diagram is similar to the case of AdS black hole \cite{Li:2020nsy}. The region above the blue line as well as below the red line corresponds to a single black hole solution which is always thermodynamically stable. The rest of the phase diagram represents three black hole solutions. Along the black curve, the free energies of small and large black holes are equal, and both these solutions coexist along this curve with equal probability. Therefore, the black line is called the coexisting curve. We observe that the free energy of the SBH phase is less than the free energy of LBH in the region between the black and red curves, and the free energy of the SBH phase is greater than the free energy of the LBH in the region between the black and blue curves. As the system having less Gibbs free energy is thermodynamically stable, we conclude that the small black hole is stable in the region between the red and black curves whereas the large black hole is thermodynamically stable in the region enclosed by the black and blue lines. However, from the ensemble point of view, one stable black hole state may turn into another stable state due to thermal fluctuations. The dynamics of such evolution of the system are described by the probabilistic Fokker-Planck equation. We will study the dynamics of the phase transition in the following section.\\

\section{Probabilistic evolution on the free energy landscape}\label{sec4}

 Now, we study the kinetics of black hole phase transition by considering black holes as thermodynamic states in the extended phase space. We have observed that large and small black hole phases can switch into each other due to the presence of thermal fluctuations. As the horizon radius $r_+$ is considered as the order parameter characterizing the black hole phases, the probability distribution of the thermodynamic state can be considered as a function of $r_+$ and time $t$.
 
  \subsection{ Fokker-Planck equation and probabilistic evolution}
\begin{figure*}[tbh]
\centering
\subfigure[ref2][]{\includegraphics[scale=0.8]{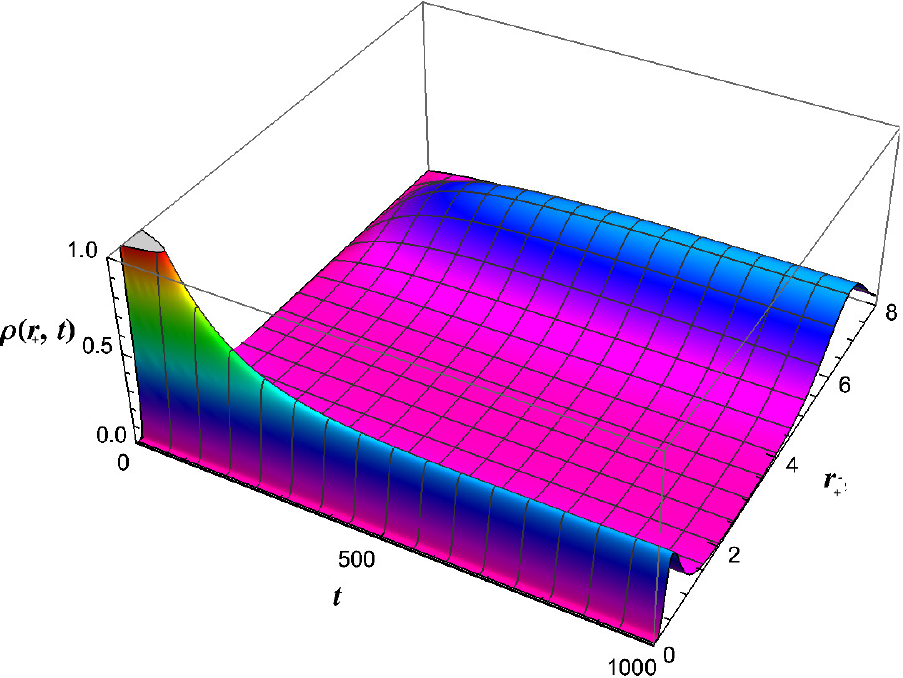}
\label{Mrbcsbh3dt1}}
\qquad
\subfigure[ref3][]{\includegraphics[scale=0.8]{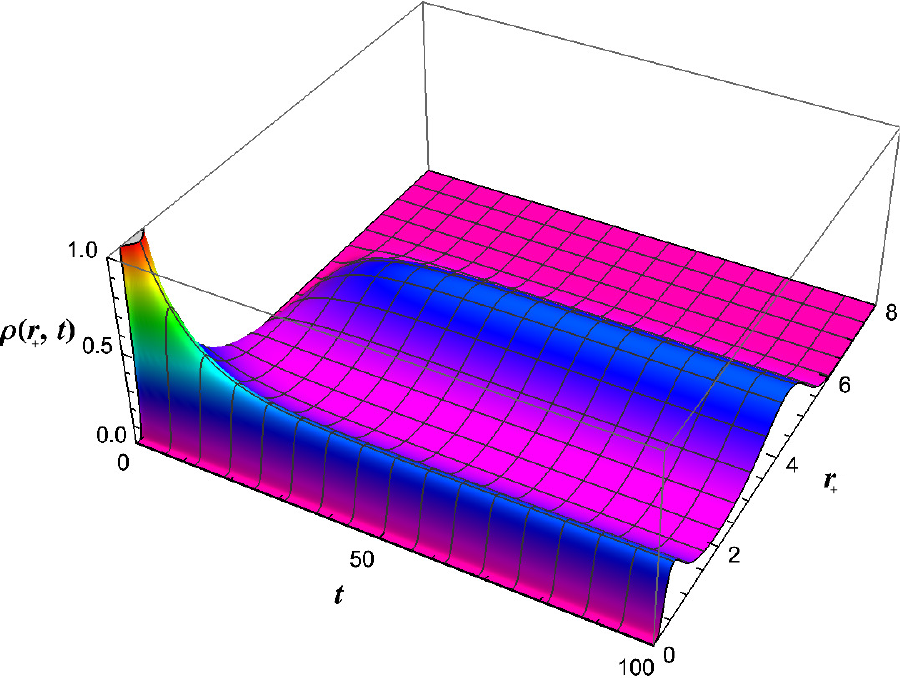}
\label{Mrbcsbh3dt2}}
\subfigure[ref2][]{\includegraphics[scale=0.8]{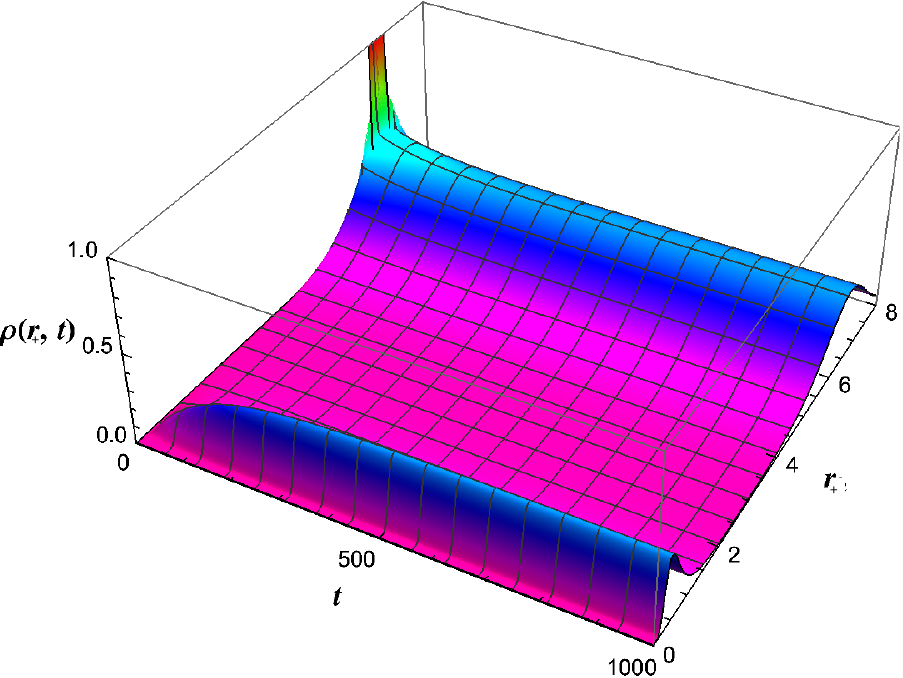}
\label{Mrbclbh3dt1}}
\qquad
\subfigure[ref3][]{\includegraphics[scale=0.8]{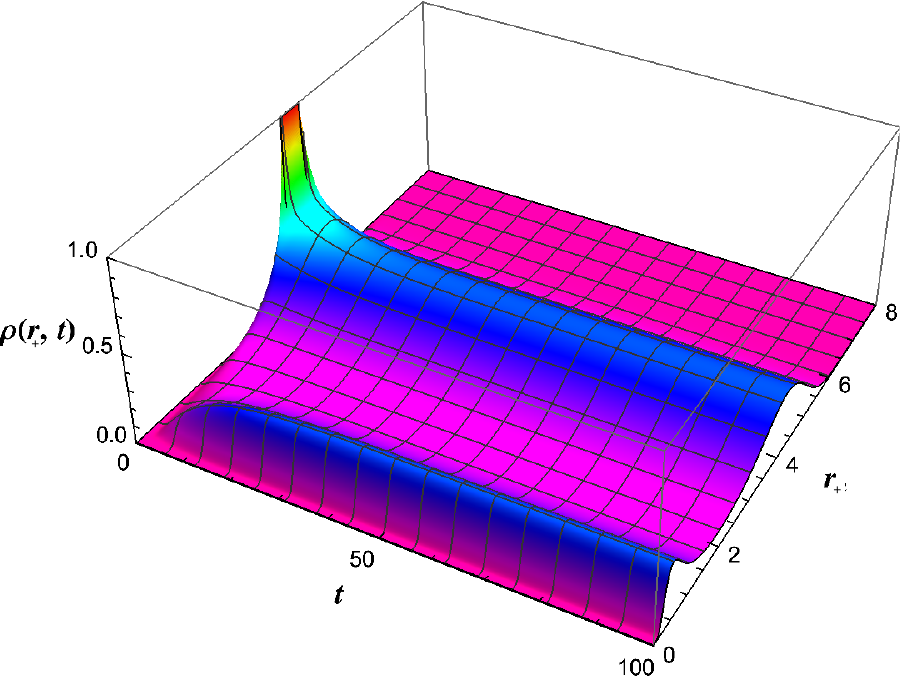}
\label{Mrbclbh3dt2}}
\caption{ A plot of probability distribution as a function of horizon radius and time. The initial black hole phase in (a) and (b) is SBH, and in (c) and (d) is LBH. The coexistent temperatures are $T_{\text{trans}}=0.4$ in left panel) and  $T_{\text{trans}}=0.5$ in right panel) with $c_1=2, q=1, m=1, c_0=1, c_2=5$ and $k=1$.}
\label{M_time_evolution}
\end{figure*}

  We denote the probabilistic distribution function of black hole states by $\rho(r_+,t)$. The evolution of the distribution function is governed by the probabilistic Fokker-Planck equation given by \citep{PhysRevE.67.041905, Lee2002FirstpassageTD},
 \begin{equation}
 \frac{\partial \rho(r_+,t)}{\partial t}= D \frac{\partial}{\partial r}\Bigg\{e^{-\beta G(r_+)}\frac{\partial}{\partial r}\Big[e^{\beta G(r_+)}\rho(r_+,t)\Big]\Bigg\}.
 \end{equation}
 Here, $D$ is the diffusion coefficient given by $D=kT/\xi$, with $k$, $\xi$ denoting the Boltzmann constant and dissipation coefficient respectively. Also, the quantity $\beta= 1/(kT)$ is the inverse temperature of the system. For convenience, we take $k=\xi=1$ without the loss of generality. To solve the Fokker-Planck equation, we need to impose two boundary conditions. The first one is the reflecting boundary condition which preserves the normalization of the probability distribution. At $r_+=r_0$, we set,
 \begin{equation*}
 e^{-\beta G(r_+)}\frac{\partial}{\partial r}[e^{\beta G(r_+)}\rho(r_+,t)]\Bigg|_{r_+=r_0}=0,
 \end{equation*}
 which can be rewritten as, 
 \begin{equation}
 \beta G'(r_+)\rho(r_+,t)+\rho'(r_+,t)\Bigg|_{r_+=r_0}=0,
 \end{equation}
 where the prime denotes the derivative with respect to the order parameter $r_+$. The second boundary condition sets the value of the distribution function to zero,
 \begin{equation}
 \rho(r_0,t)=0.
 \end{equation}
 In the following analysis, we chose the reflecting boundary condition at $r_0=0$ and $r_0=\infty$. The initial distribution is taken to be a Gaussian wave packet located at $r_i$, which is a good approximation of $\delta$- distribution.
 \begin{equation}
 \rho(r_+,0)=\frac{1}{\sqrt{\pi}a} e^{-\frac{(r-r_i)^2}{a^2}},
 \end{equation}
 where the parameter $a$ is a constant that determines the initial width of the wave packet. Note that the initial distribution is well normalized, and as a consequence of the reflection boundary condition, $\rho(r_+,t)$ will remain normalized during the evolution. The parameter $r_i$ denotes the radius of the initial black hole state. We may choose either $r_i=r_s$ representing SBH as the initial state or $r_i=r_l$ for LBH. Suppose we chose $r_i=r_l$ at $t=0$. As the distribution evolves, we observe a non-zero probability distribution for both SBH and LBH states. This indicates the phase transition between LBH and SBH phases due to thermal fluctuations. The time evolution of $\rho(r_+,t)$ is plotted in Fig. (\ref{M_time_evolution}).

In Fig. \ref{Mrbcsbh3dt1} and \ref{Mrbcsbh3dt2}, we set $r_i=r_s$, and studied the evolution of the distribution for two values of the transition temperatures ($T_{\text{trans}} = 0.4, 0.5$). Initially, the probabilistic distribution is peaked at $r_+ = r_s$. The probabilistic distribution becomes quasi-stationary in a short time with peaks at small and large black hole states. After some time, the height of $\rho(r_+,t)$ at $r_+=r_s$ decreases while another peak starts to develop at $r_+=r_l$. This implies the leakage of SBH state to LBH state. Further, as $t\rightarrow\infty$, $\rho(r_+,t)$ becomes a stationary state. A similar behavior is observed if we take the initial state to be a large black hole. The corresponding evolution for two different temperatures is depicted in Fig. \ref{Mrbclbh3dt1}, and \ref{Mrbclbh3dt2}.\\

\begin{figure*}[tbh]
\centering
\subfigure[ref2][]{\includegraphics[scale=0.8]{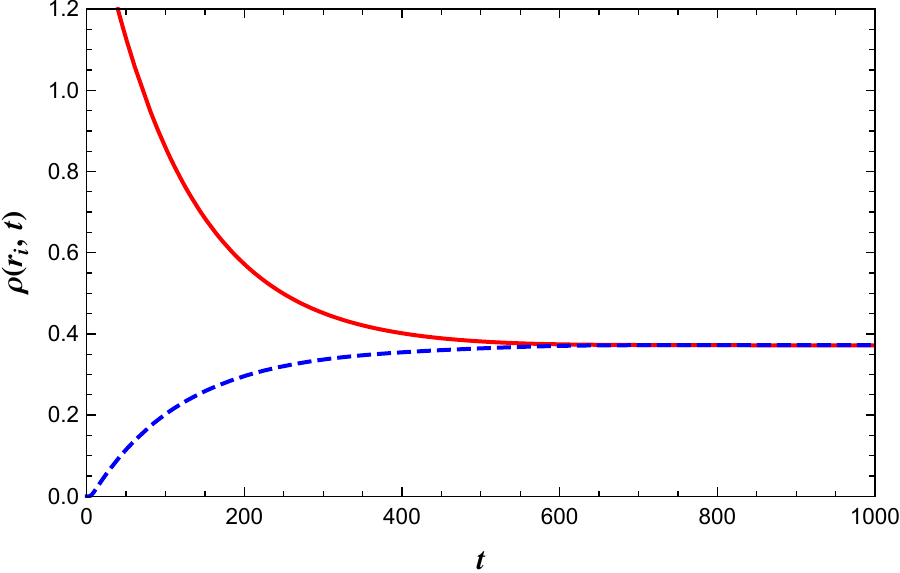}
\label{Mrbcsbh2dt1}}
\qquad
\subfigure[ref3][]{\includegraphics[scale=0.8]{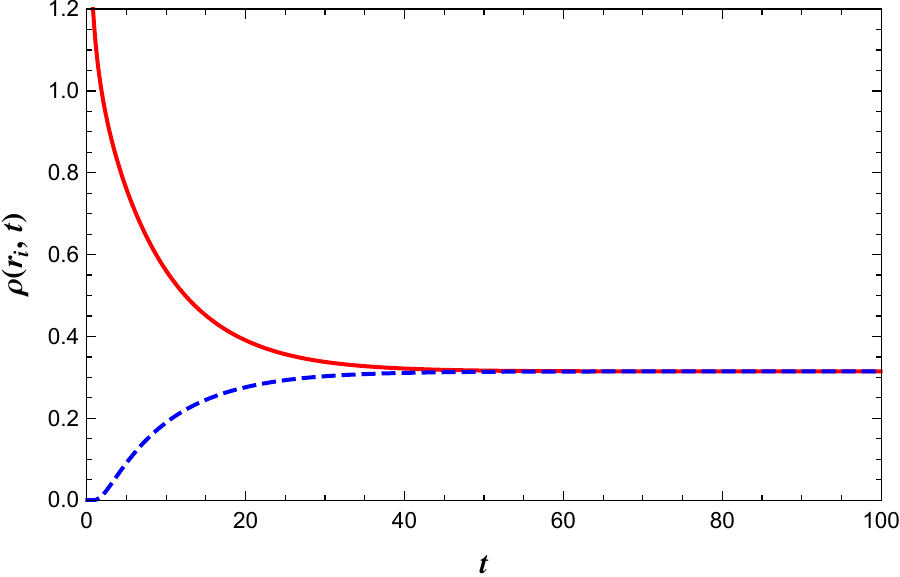}
\label{Mrbcsbh2dt2}}
\caption{ The evolution of probability distribution function with the initial Gaussian wave packet at SBH is plotted for two values of coexisting temperatures. The red solid and blue dashed lines represent $\rho (r_{s},t)$ and $\rho (r_{l},t)$, respectively. The coexistent temperatures are (a) $T_{\text{trans}}=0.4$. (b) $T_{\text{trans}}=0.5$, with $c_1=2, q=1, m=1, c_0=1, c_2=5$ and $k=1$.}
\label{time_evolution_2d}
\end{figure*}

The above analysis can be made more apparent by comparing the behavior of $\rho(r_s,t)$ and $\rho(r_l,t)$ as time progresses, where $\rho(r_s,t)$, $\rho(r_l,t)$ represent the probability distribution of SBH and LBH states respectively. In Fig. \ref{time_evolution_2d}, the evolution of probability distribution for both black hole states is depicted for two different transition temperatures. In both cases, we have taken the initial black hole state to be SBH. In other words, at $t=0$, $\rho(r_s)$ takes a finite value but $\rho(r_l)$ vanishes. Fig. \ref{Mrbcsbh2dt1} describes the evolution of both probability distributions for a transition temperature $T_{\text{trans}}=0.4$. Now, as time increases, the height of $\rho(r_s,t)$ decreases while that of $\rho(r_l,t)$ increases, indicating the leakage of black hole states from SBH to LBH. For large values of time, both $\rho(r_s,t)$ and $\rho(r_l,t)$ approaches a final stationary state where $\rho(r_s)=\rho(r_l)$. In Fig. \ref{Mrbcsbh2dt2}, the same evolution is drawn but for a higher transition temperature, $T_{\text{trans}}=0.5$. From the figure, we observe that for larger transition temperatures, both distributions reach the stationary state more rapidly. A similar analysis can be carried out by taking other black hole states as the initial configuration giving the same results. \\

In the coming session, we employ the first passage process to discuss the kinetics of black hole phase transitions in massive gravity theory. The first passage time is an important quantity as the average timescale for a stochastic process to occur for the first time is characterized by mean first passage time.

\subsection{First passage time}
\begin{figure*}[tbh]
\centering
\subfigure[ref2][]{\includegraphics[scale=0.8]{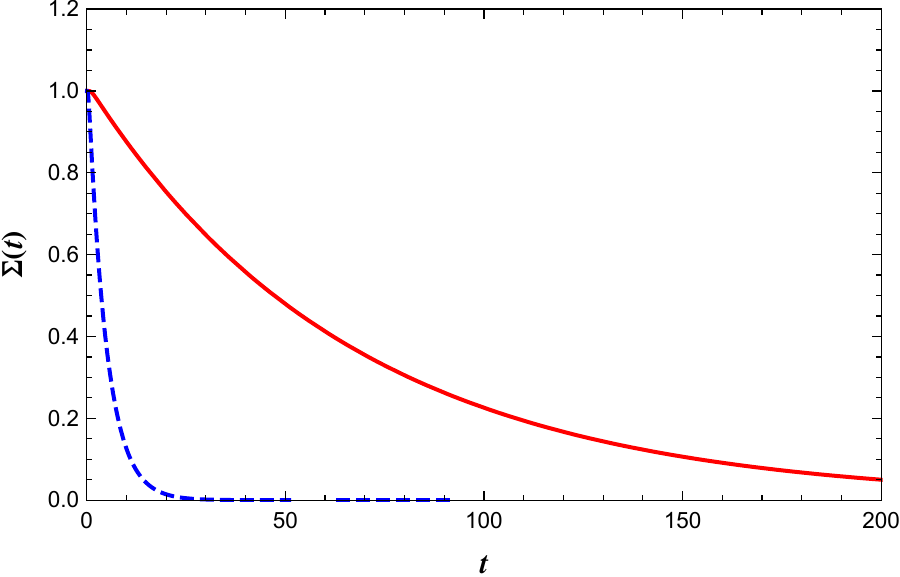}
\label{probsbh}}
\qquad
\subfigure[ref3][]{\includegraphics[scale=0.8]{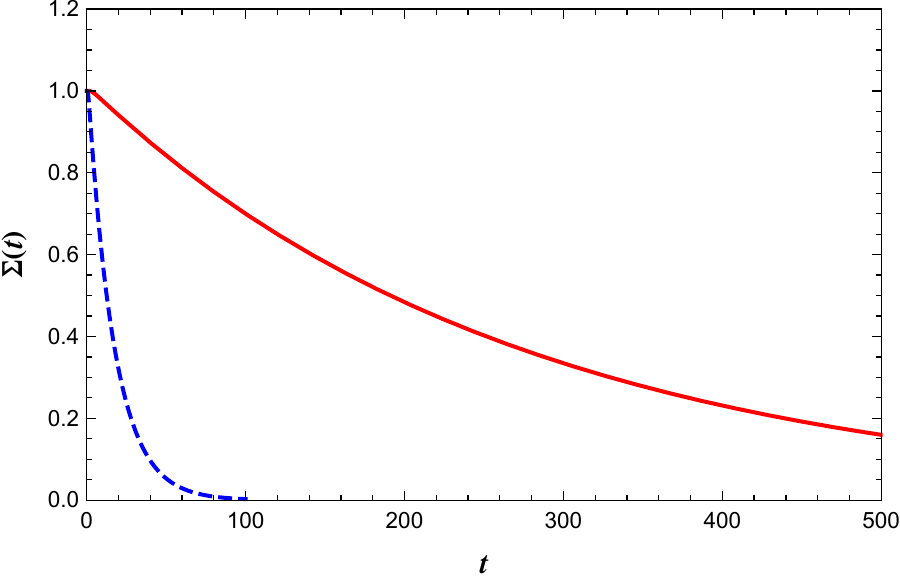}
\label{problbh}}
\caption{The time evolution of $\Sigma (t)$; (a) the state remains at the initial state SBH, and  (b) the system remains at LBH. Red solid and blue dashed curves are for the coexistent temperatures $T_{\text{trans}}=0.4$ and  $T_{\text{trans}}=0.5$, respectively, with $c_1=2, q=1, m=1, c_0=1, c_2=5$ and $k=1$.}
\label{prob}
\end{figure*}

\begin{figure*}[tbh]
\centering
\subfigure[ref2][]{\includegraphics[scale=0.7]{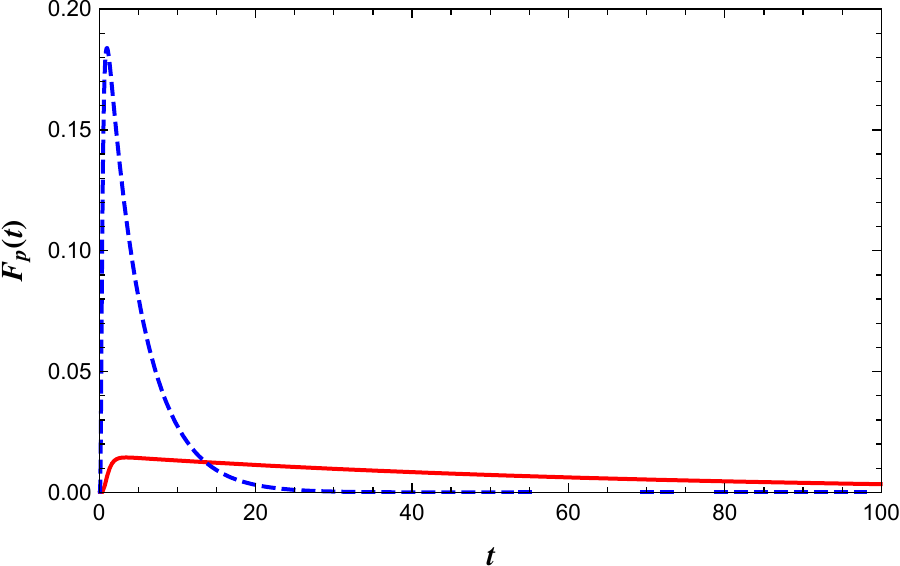}
\label{firstpasssbh}}
\qquad
\subfigure[ref3][]{\includegraphics[scale=0.7]{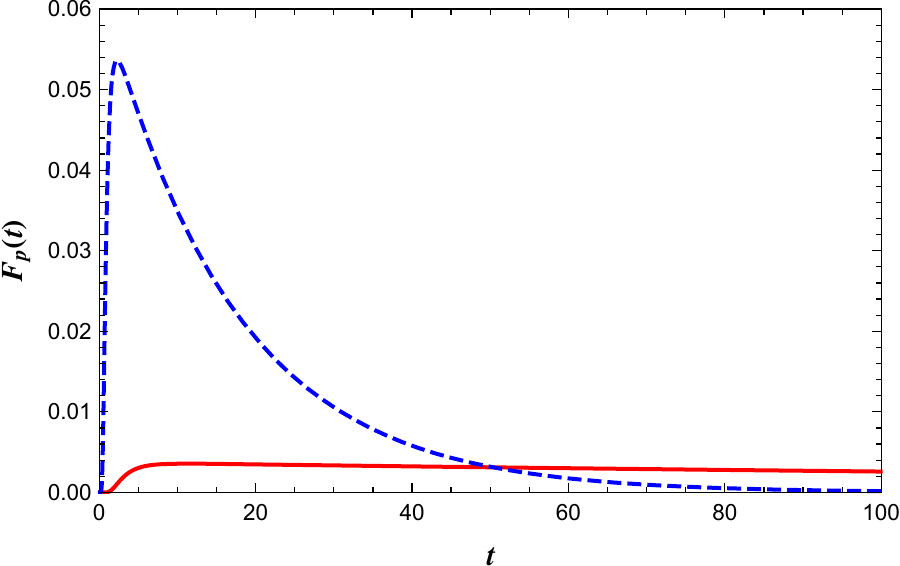}
\label{firstpasslbh}}
\caption{A plot of $F_p(t)$ as the function of time: (a) from SBH to LBH, and (b) from LBH to SBH. Solid red and dashed blue lines represent the coexistent temperatures $T_{\text{trans}}=0.4$ and  $T_{\text{trans}}=0.5$, respectively.  Here, $c_1=2, q=1, m=1, c_0=1, c_2=5$ and $k=1$.}
\label{firstpass}
\end{figure*}

In transition state theory, the first passage time is defined as the time required for a stable thermodynamic phase to reach an unstable transition state for the first time due to thermal fluctuations in the system. For black hole system, we define the first passage time as the time required for a stable black hole state (SBH) to reach the unstable transition state, represented by the peak of the Gibbs free energy. Now, the probability that the present state of SBH that has not made a first passage by time $t$ is defined as \citep{Wei:2020rcd},
\begin{equation}
\Sigma(t) = \int_0^{r_m} \rho(r_+,t)dr_+,
\end{equation}
where $r_m$ is the radius of the intermediate black hole state obtained from Eq. (\ref{radius}). As time progress, the probability that the system remains at SBH decreases and approaches zero as $t\rightarrow \infty$, i.e., $\Sigma(t)_{t \rightarrow \infty}=0$. This is due to the fact that the normalization of the probability distribution is preserved in this case. Similarly, one can start with an LBH state and define the probability distribution that the initial LBH state has not made a first passage by time $t$ as,
\begin{equation}
\Sigma(t) = \int_{r_m}^{\infty} \rho(r_+,t)dr_+,
\end{equation}
The time evolution of $\Sigma(t)$ for both SBH and LBH for two values of transition temperatures are plotted in Fig. \ref{prob}. The probability distribution is clearly seen to vanish for long-time evolutions. Further, the probability distribution decreases faster for transitions at higher temperatures.\\
\\
 Now, the first passage time is the probability that a small black hole state passes through the intermediate state for the first time in the interval ($t,t+dt$). The distribution of first passage time is given by,
\begin{equation}
F_p(t) = -\frac{d\Sigma(t)}{dt}.
\end{equation}
Substituting for $\Sigma(t)$ and using the Fokker-Planck's equation along with the reflecting boundary condition at $r_+=0$ and absorbing boundary condition at $r_+=r_m$, we get the following expression for the first passage time. 
\begin{equation}
F_p(t) = -D\frac{\partial}{\partial t}\rho(r_+,t)\Bigg|_{r=r_m}.
\end{equation}
The distribution of first passage time for both SBH to LBH and LBH to SBH phase transitions are plotted for different transition temperatures (Fig. \ref{firstpass}). In Fig. \ref{firstpasssbh}, the initial distributions are Gaussian wave packets located at the small black hole state. The single peak in the first passage time within a short period indicates that a considerable fraction of the first passage events have occurred before the distribution attains its exponential decay form. As time increases the peak becomes sharper. The curves corresponding to different transition temperatures show similar behavior of the first passage time. Further, when the initial state is SBH, the location of the peak moves to the left (lower value of time) when the transition occurs at larger transition temperatures. These characteristics can be justified by looking at the behavior of the barrier height between the small and intermediate black hole states as a function of temperature. As mentioned in \ref{sec3}, $G(r_m)-G(r_s)$ decreases as temperature increases. Therefore, the small black hole state can cross the barrier to reach the intermediate state easily at higher transition temperatures. However, if the initial distribution is peaked at a large black hole state, the location of the peak moves to the right (higher values of time) when the transition occurs at larger transition temperatures Fig. \ref{firstpasslbh}. Similar to the previous case, this behavior is explained as follows: As the barrier height $G(r_m)-G(r_l)$ increases with the temperature, the large black hole state takes more time to cross the barrier under thermal fluctuations. These results are qualitatively similar to the case of dynamics of phase transitions charged AdS black holes presented in \cite{Li:2020nsy}.

In the following, we investigate the role of mass and the topological parameter on the results discussed above on the dynamic phase transition of black holes in massive gravity theory.

\section{The effect of mass and topology}\label{sec5}

Note that the numerical results presented so far in the investigation of the dynamics of black hole phase transition in dRGT non-linear massive gravity theory consider only certain values of the parameters in the system. It remains to study the behavior of time evolution of the probability distribution and the first passage time by varying those parameters. However, as the theory possesses too many variables, we will be considering the effect of mass and topology only.\\

\begin{figure}[H]
\centering
\subfigure[ref2][]{\includegraphics[scale=0.7]{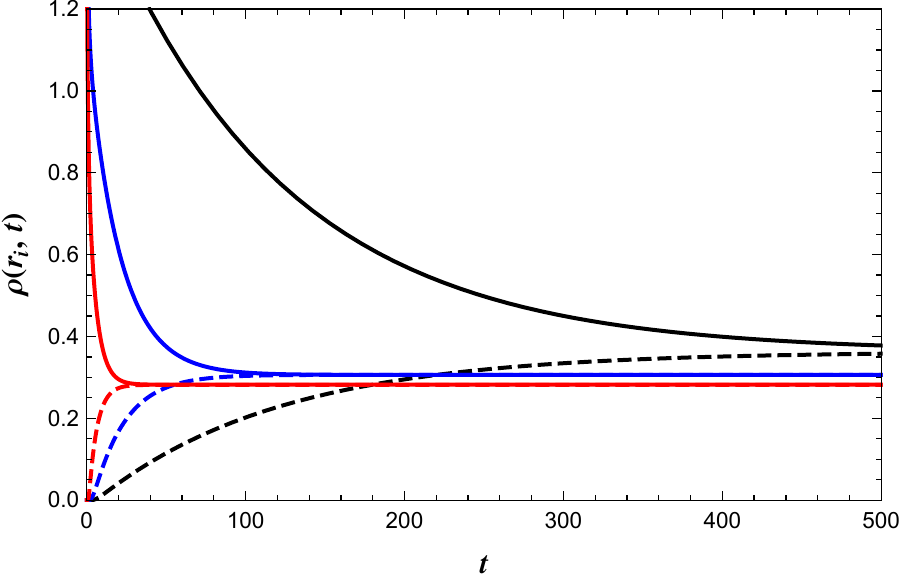}
\label{Msbhk}}
\qquad
\subfigure[ref3][]{\includegraphics[scale=0.7]{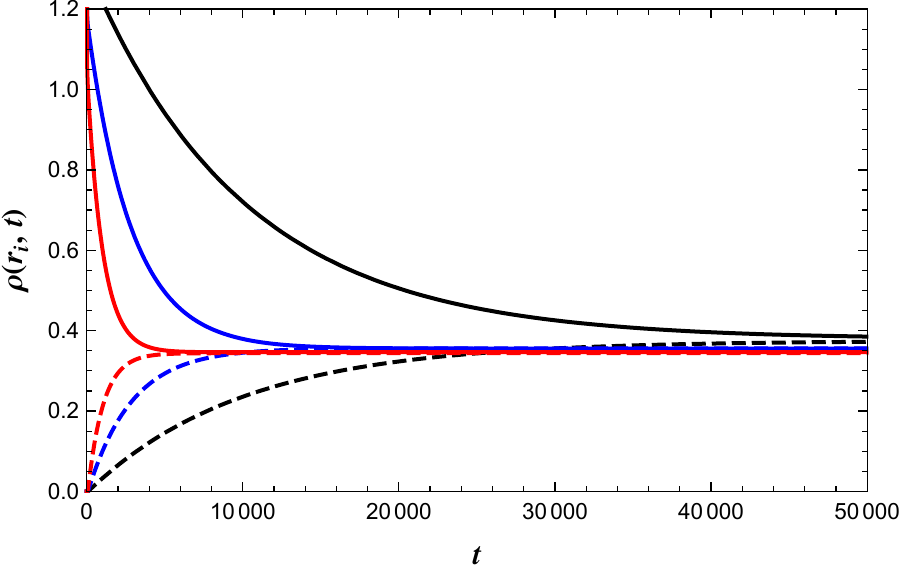}
\label{Msbhm}}
\caption{\small{The effect of the parameters on the time evolution of the probability distribution $\rho (r_+,t)$. The solid and dashed curves correspond to the functions $\rho (r_{s},t)$ and $\rho (r_{l},t)$, respectively. The initial Gaussian wave packet is located at the SBH state. (a) The effect of topology. Black, blue, and red correspond to $k=1,0,-1$, respectively. The coexistent temperature is $T_{\text{trans}}=0.4$. (b) The effect of mass parameter $m$. Red, blue, and black correspond to $m=0,0.07,0.1$, respectively. The coexistent temperature is $T_{\text{trans}}=0.03$. The other parameters in these two plots are $c_1=2, q=1, m=1, c_0=1, c_2=5$, and $k=1$.}}
\label{M_time_evolution_2d}
\end{figure}

In Fig. \ref{M_time_evolution_2d}, the evolution of the probability distribution function is plotted for different topologies and masses. Here, we have considered planar ($\kappa=0$), hyperbolic ($\kappa=-1$), and spherical ($\kappa=1$) topologies. For the sake of demonstration, the initial black hole is taken as the SBH state and the transition temperature is $T_{\text{trans}}= 0.4$ (Fig. \ref{Msbhk}). One can see from the figure that the height of $\rho(r_s,t)$ decreases rapidly for $\kappa=-1$, indicating a quick leakage of the black hole state from SBH to LBH. In the case of planar topology, the transition is less rapid and in the spherical case, the evolution is slow. This means that out of three, the spherical black holes attain the stationary state only after a long evolution. Further, the value of the probability distribution at the stationary state ($\rho(r_s) = \rho(r_l)$) is highest for the spherical case compared to the planar and hyperbolic cases. The effect of the mass parameter shows similar behavior. As $m$ increases, the system attains its final stationary state slowly, i.e., as the mass increases the small black hole state leaks towards the large black hole state slowly (Fig \ref{Msbhm}).\\

\begin{figure}[H]
\centering
\subfigure[ref2][]{\includegraphics[scale=0.7]{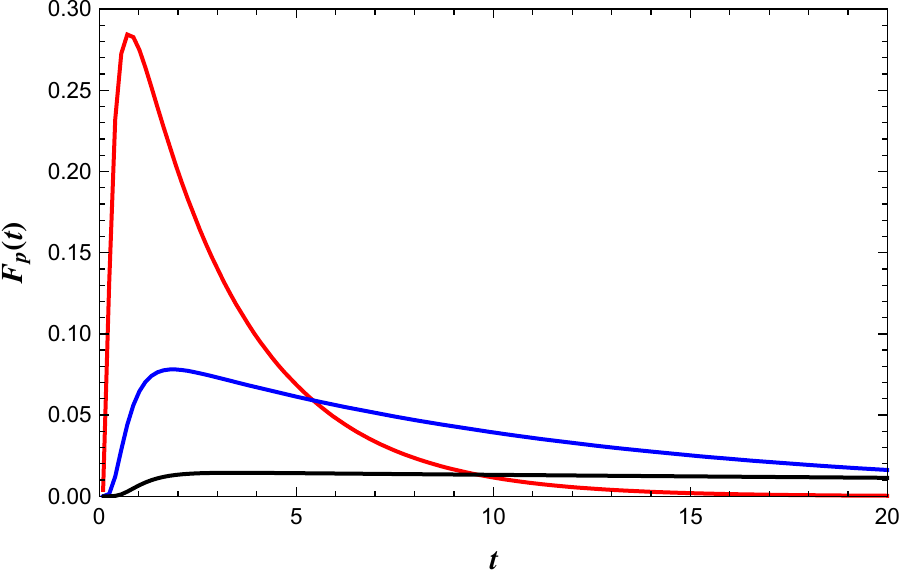}
\label{Msbhfirstpassk}}
\qquad
\subfigure[ref3][]{\includegraphics[scale=0.7]{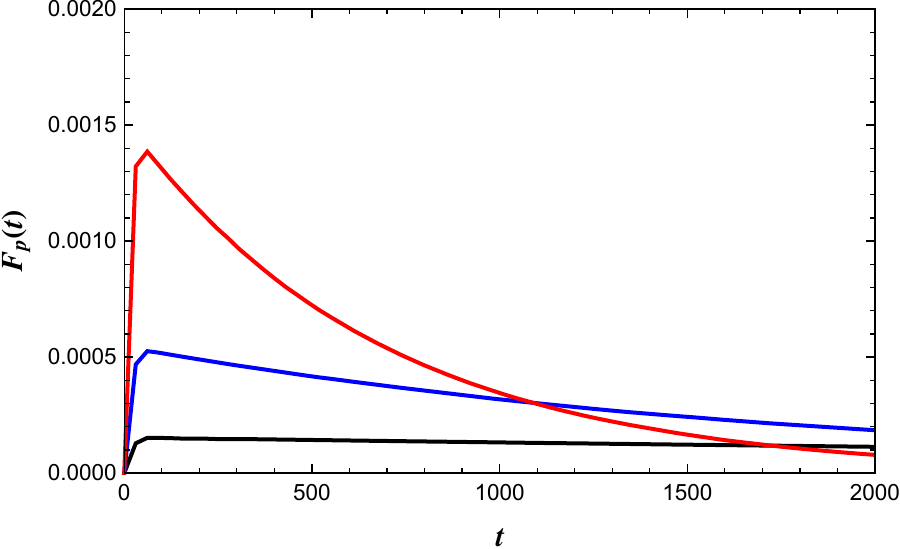}
\label{Msbhfirstpassm}}
\caption{\small{The effect of the spacetime parameters on the probability distribution of the first passage time $F_p(t)$ is shown. We consider the case for  SBH (initial) state to LBH state. (a) The effect of topology. Black, blue, and red solid lines are for $k=1,0,-1$, respectively. The coexistent temperature is $T_{\text{trans}}=0.4$. (b) The effect of mass parameter $m$. Red, blue, and black correspond to $m=0,0.07,0.1$, respectively. The coexistent temperature is $T_{\text{trans}}=0.03$. The other parameters in these two plots are $c_1=2, q=1, m=1, c_0=1, c_2=5$, and $k=1$.}}
\label{firstpass_m}
\end{figure}

Next, we address the effect of topology and mass on the probability distribution of first passage time. In Fig. \ref{firstpass_m}., we have considered an initial SBH state with the topologies $\kappa =0, -1$ and $1$. For a given transition temperature, theSBH state reaches the intermediate state in a small duration of time for hyperbolic topology. In the case of spherical topology, the corresponding time is maximum Fig. \ref{Msbhfirstpassk}. Further, we observe that the time taken to reach the intermediate state, starting from a SBH increases as mass increases (Fig. \ref{Msbhfirstpassm}).\\

\section{Discussions}
\label{sec6}
In this paper, we have studied the dynamics of black hole phase transitions in dRGT non-linear massive gravity theory using the free energy landscape. Thermodynamic characterization and different black hole phases are discussed by considering the black hole radius as the order parameter. Emergent phases of small and large black holes as well as the coexisting curve between these states are shown. The switching of one black hole phase to another due to thermal fluctuations is addressed in terms of first passage time. Further, we have solved the Fokker-Planck equation numerically and the results are explained. The results we have presented are qualitatively similar to the findings of \citep{Li:2020nsy}. Finally, we have discussed the effect of mass parameters and the topology on the evolution of black hole phase transition between small and large black hole states. We believe that the dependance of switching time from small black holes to large black hole states with mass and the topology is connected to the stability of black holes and will be addressed in future works. Also, we would like to revisit the properties of black hole phase transitions by incorporating Hawking radiation.\\
\\

\acknowledgments
TKS thanks Aparna L. R. for the support and suggestions.\\
\\
\bibliography{MassiveFP}

\end{document}